\documentclass[12pt]{article}
\usepackage{authblk,cite,amssymb,amsmath,xcolor,graphics,epstopdf,graphicx}
\begin{document}
\title{6D Standing Wave Braneworld with Ghost Scalar Fields}
\author{{\bf Otari Sakhelashvili}\\
Javakhishvili Tbilisi State University\\
3 Chavchavadze Avenue, Tbilisi 0179, Georgia\\
{\sl E-mail: otosaxel@gmail.com}}
\maketitle

\begin{abstract}
The 6D braneworld with the phantom-like bulk scalar field is considered. We demonstrate pure gravitational localization of scalar field zero modes on the brane.

\end{abstract}

%%%%%%%%%%%%%%%%%%%%%%%%%%%%%%%%%%%%%%%%%%%%%%%%%%%%%%%%%%%%%%%%%%%%%%%%%%%%%%%%%%%%%%%%%%%%%%%%%%

\section{Introduction}

Braneworld models involving large extra dimensions \cite{Hi,brane} have been very useful in addressing several open questions in high energy physics (see \cite{reviews} for reviews). Most of the braneworlds are realized as time-independent field configurations. However, there have appeared several braneworld models that assumed time-dependent metrics and fields \cite{S,Wave,6D}.

In this paper we study the non-stationary 6D braneworld generated by standing gravitational waves coupled to a phantom-like bulk scalar field. The model is generalization of 5D standing waves scenario of \cite{Wave}. The metric we use is special case of the general solution found in \cite{GMT} and slightly differs from those of the 6D models \cite{6D}.

%%%%%%%%%%%%%%%%%%%%%%%%%%%%%%%%%%%%%%%%%%%%%%%%%%%%%%%%%%%%%%%%%%%%%%%%%%%%%%%%%%%%%%%%%%%%%%%%%%

\section{The solution of Einstein equations}

We consider 6D space-time, having the signature $(+,-,-,-,-,-)$, with a non-self interacting phantom-like scalar field coupled to gravity. The action of the model is:
\begin{equation}\label{action}
S = \int d^6x \sqrt g \left( \frac{M^4}{2}R - \frac 12 g^{MN}\partial _M \phi \partial _N\phi \right)~,
\end{equation}
where $M$ is the fundamental scale, which relates to the 6D Newton constant, $G=1/(8\pi M^4)$. Capital Latin indexes numerate the coordinates of 6D space-time and we use the units where $c = \hbar = 1$. Variation of the action (\ref{action}) with respect to $g_{AB}$ leads to the 6D Einstein equations:
\begin{equation}\label{Einstein1}
R_{AB} - \frac 12 g_{AB}R = \frac {1}{M^4}T_{AB}~,
\end{equation}
where
\begin{equation}\label{ScalarT}
T_{AB} = - \partial _A\phi \partial _B\phi + \frac 12 g_{AB} \partial ^C\phi \partial _C\phi~.
\end{equation}
is the energy momentum tensor of the scalar field. Using (\ref{ScalarT}) the Einstein equations (\ref{Einstein1}) can be rewritten in the form:
\begin{equation}\label{Einstein2}
R_{AB}= -\frac {1}{M^4}\partial_A \phi \partial_B \phi~.
\end{equation}

To solve the equations (\ref{Einstein2}) we take the metric {\it ansatz}:
\begin{equation} \label{ansatz}
ds^2 = e^S dt^2 - e^u \left(dx^2 + dy^2 + dz^2\right) - dr^2 - e^{-3u} d\theta^2~,
\end{equation}
where $S=S(r)$, $u=u(t,|r|)$ are some functions, and suppose that phantom-like scalar field depends only on the time and on the modulus of one extra dimension coordinate $r$. By the metric (\ref{ansatz}) we want to describe geometry of the brane placed at the origin of the large extra dimension, $r$, when $x$, $y$ and $z$ denote coordinates along the brane. The sixth coordinate $\theta$ can be assumed to be compact, curled up to the unobservable sizes for the present energies.

Advantage of our 6D {\it ansatz} (\ref{ansatz}) is that it looks symmetric in the brane coordinates $x$, $y$ and $z$, while in 5D model \cite{Wave} the brane 3-plain is non-symmetrically warped. Also (\ref{ansatz}) is simpler than the metric of the models \cite{6D}, since it contains only one singular point, $r = 0$, where the brane is placed to smooth the singularity.

The system of Einstein equations (\ref{Einstein2}) for the {\it ansatz} (\ref{ansatz}) takes the form:
\begin{eqnarray} \label{system_1}
3\dot{u}^2 - \frac 12 S''e^S - \frac 14 S'^2 e^S &=& \frac {1}{M^4}\dot{\phi}^2~, \nonumber\\
3\dot{u}u' &=& \frac {1}{M^4}\dot{\phi}\phi'~, \nonumber\\
\frac 12 S'' + \frac 14 S'^2 + 3u'^2 &=& \frac {1}{M^4} \phi'^2~,\\
-2\ddot{u} + S'e^S u'+2u''e^S &=& 0~, \nonumber
\end{eqnarray}
where overdots and primes mean derivatives with respect to $t$ and $|r|$, respectively. If we assume that the metric field, $u$, is proportional to the bulk phantom field, $\phi$, and separate the variables:
\begin{equation} \label{u}
u(t,|r|) = \frac{1}{\sqrt {3M^4}}~\phi(t,|r|) = \sin(\omega t) f(|r|)~,
\end{equation}
the system (\ref{system_1}) reduces to:
\begin{eqnarray} \label{system_2}
S'' + \frac 12 S'^2 &=& 0~, \nonumber\\
2\omega^2 f + S' e^S f' + 2f''e^S &=& 0~.
\end{eqnarray}
Solution of the first equation of (\ref{system_2}) with the boundary condition:
\begin{equation}
S(0) = 0~,
\end{equation}
(leading to the Minkowski metric at $r=0$) is;
\begin{equation}\label{S_sol}
S = \ln \left(1 + \frac {|r|}{a} \right)^2~,
\end{equation}
where $a$ is some constant. Using (\ref{S_sol}) the second equation of (\ref{system_2}) takes the form:
\begin{equation}
f'' + \frac{1}{a + |r|} f' + \frac{a^2\omega^2}{(a + |r|)^2}f = 0~.
\end{equation}
Solution to this equation leading to the Minkowski metric at $r=0$ is:
\begin{equation}
f(|r|) = C\sin \left( a\omega \ln\left[1+\frac{|r|}{a}\right]\right) ~,
\end{equation}
where $C$ is a constant.

Finally the metric function (\ref{u}) obtains the form:
\begin{equation}
u(t,|r|) = C\sin (\omega t) \sin \left( a\omega \ln\left[1+\frac{|r|}{a}\right]\right)~,
\end{equation}
and the solution of Einstein equations for our {\it ansatz} (\ref{ansatz}) is:
\begin{eqnarray} \label{ansatz_solv}
ds^2 &=& \left( 1 + |r|/a \right)^2dt^2 - e^{C\sin (\omega t) \sin \left( a\omega \ln[1+|r|/a]\right)} \left( dx^2 + dy^2 + dz^2 \right) - \nonumber \\
&-& dr^2 -e^{-3C\sin (\omega t) \sin \left( a\omega \ln[1+|r|/a]\right)} d\theta^2~.
\end{eqnarray}

%%%%%%%%%%%%%%%%%%%%%%%%%%%%%%%%%%%%%%%%%%%%%%%%%%%%%%%%%%%%%%%%%%%%%%%%%%%%%%%%%%%%%%%%%%%%%%%%%%%%%%%%

\section {Time averages}

When the frequency of standing gravitational waves, $\omega$, is much larger than frequencies associated with the energies of the particles on the brane,
\begin{equation}
\omega \gg E~,
\end{equation}
we can perform time averaging of oscillating exponents in the equations of matter fields. Using the known formula for averages:
\begin{equation} \label{av_expsin}
\left\langle e^{D \sin (x)}\right\rangle = I_0(D)~,
\end{equation}
where $I_0$ is modified Bessel function of zero order, we can find the time average of our {\it ansatz} (\ref{ansatz}):
\begin{eqnarray} \label{av_ansatz}
\left\langle ds^2 \right\rangle &=& ( 1 + |r|/a )^2 dt^2 - I_0(C\sin \left( a\omega \ln[1+|r|/a]\right)) \left( dx^2 + dy^2 + dz^2 \right) - \nonumber \\
&-& dr^2 -I_0(-3C\sin \left( a\omega \ln[1+|r|/a]\right)) d\theta^2
\end{eqnarray}

%%%%%%%%%%%%%%%%%%%%%%%%%%%%%%%%%%%%%%%%%%%%%%%%%%%%%%%%%%%%%%%%%%%%%%%%%%%%%%%%%%%%%%%%%%%%%%%%%%%%%%%

\section{Localization of scalar fields}

Let us consider the localization problem for the massless scalar field, $\Phi$, defined by the 6D action:
\begin{equation}\label{Action}
S_\Phi = \frac 12 \int d^{6}x \sqrt{g}g^{MN} \partial_M\Phi\partial_N\Phi~.
\end{equation}
The corresponding Klein-Gordon equation is:
\begin{equation}\label{ScalFieldEqn}
\frac{1}{\sqrt g}~\partial_M \left( \sqrt g g^{MN}\partial_N \Phi \right) = 0~.
\end{equation}
We look for the solution of this equation in the form:
\begin{equation}\label{fsi}
\Phi(x^A)=\psi(t,x,y,z)\sum_l\nu_l(r)e^{il\theta}
\end{equation}
Let us consider the $S$-wave solution ($l=0$) and nothing depends on the extra dimension angle $\theta$. After inserting of (\ref{fsi}) into the action (\ref{Action}) and performing of time averaging we find:
\begin{equation}\label{Action_last}
S_\Phi = \frac 12 \int d^6x\left\{\frac{\nu^2 (\partial_t\psi)^2}{1 + r/a} - \left(1+\frac{r}{a}\right)\nu'^{2}\psi^{2} - \left(1+\frac{r}{a}\right)\nu^{2} \left\langle e^u\right\rangle \left[(\partial_x\psi)^2 + (\partial_y\psi)^2 + (\partial_z\psi)^2 \right] \right\}.
\end{equation}
In general, to have a field localized on a brane 'coupling' constants appearing after integration of the Lagrangian over the extra coordinates must be non-vanishing and finite. So normalizable zero modes of our scalar field, $\Phi$, on the brane will be exists if the action (\ref{Action_last}) is integrable over $r$, i.e. the functions $(1 + r/a)\nu'^2$, $(1 + r/a) \left\langle e^{u}\right\rangle \nu^2$ and $\frac{1}{1+r/a}\nu^2$ are integrable.

On the background (\ref{ansatz_solv}) the differential operator in the Klein-Gordon equation (\ref{ScalFieldEqn}), after time averaging, gets the form:
\begin{equation}\label{OP_1}
\widehat{L} = \frac{1}{(1+r/a)}\frac{\partial^2}{\partial t^2} - \left(1+\frac{r}{a}\right)\left\langle e^{-u}\right\rangle \left(\frac{\partial^2}{\partial x^2} + \frac{\partial^2}{\partial y^2} + \frac{\partial^2}{\partial z^2}\right) -  \frac{\partial}{\partial r}\left(1+\frac{r}{a}\right)\frac{\partial}{\partial r} - \left\langle e^{3u}\right\rangle \frac{\partial^2}{\partial \theta^2}~.
\end{equation}
Applying this operator for the $S$-wave part of the scalar field wavefunction (\ref{fsi}) we find:
\begin{equation} \label{system_KG}
\left[\left(1 + \frac {r}{a}\right)\nu(r)'\right]' + \left[\frac{1}{(1+ r/a)}E^{2} -\left(1+\frac{r}{a}\right)\left\langle e^{u}\right\rangle \left(p_x^2 + p_y^2 + p_z^2 \right)\right]\nu(r) = 0~.
\end{equation}
Transformation of the variable,
\begin{equation}
\nu(r)=\frac{\phi}{\sqrt{1 + r/a}}~,
\end{equation}
brings (\ref{system_KG}) to the Schrodinger-like form:
\begin{equation} \label{SR_LIKE}
\phi''-U(r)\phi = 0~.
\end{equation}
Here the potential function is:
\begin{equation} \label{SR_LIKE}
U(r) = \left\langle e^{-u}\right\rangle  (p_x^2 + p_y^2 + p_z^2) -\frac{a^2E^2}{(a+r)^2} -\frac{1}{4(a+r)^2}~,
\end{equation}
Figure 1 displace the shape of $U(r)$ for the positive $r$-s.
\\\\
If we would assume:
\begin{equation}
\left\langle e^{-u}\right\rangle \left(p_x^2 + p_y^2 + p_z^2\right)\approx P^{2}~.
\end{equation}

%%%%%%%%%%%%%%%%%%%%%%%%%%%%%%%%%%%%%%%%%%%%%%%%
\begin{figure}[!htp]
\begin{center}
\includegraphics[width=0.5\textwidth]{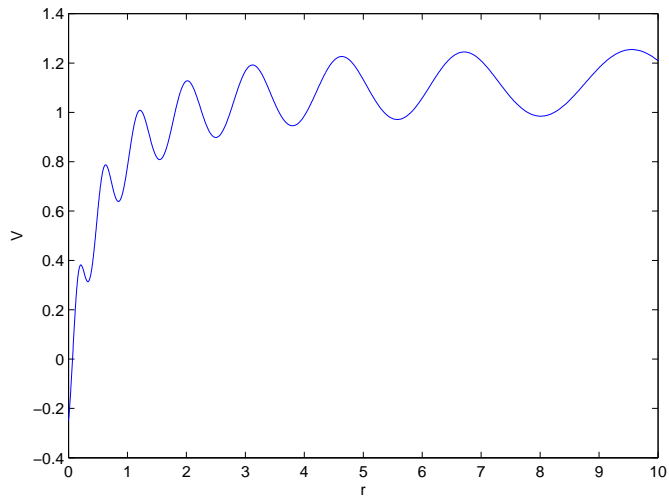}
\caption{Shape of the effective bulk potential $U(r)$ for scalar zero mode.}
\label{fig:pot}
\end{center}
\end{figure}
%%%%%%%%%%%%%%%%%%%%%%%%%%%%%%%%%%%%%%%%%%%%%%%

Close to the brane $r \ll a$ and the equation (\ref{SR_LIKE}) takes the form:
\begin{equation}\label{Eq-phi}
\phi''-\left(P^2 - E^2 - \frac{1}{4a^2}\right)\phi = 0~.
\end{equation}
Close to the origin  $r \to 0$, i.e. on the brane, this equation gives the standard dispersion relation,
\begin{equation}
E^{2}=p_x^{2}+p_y^{2}+p_z^{2}~,
\end{equation}
for zero mode particle. It is easy to find that the equation (\ref{Eq-phi}) has the plane wave solution:
\begin{equation} \label{sol-0}
\phi|_{r\to 0} = C_1e^{ir/2a} + C_2e^{-ir/2a}~,
\end{equation}
where $C_1$ and $C_2$ are constants.

Fare from the brane, $r\to \infty$, the equation (\ref{SR_LIKE}) reduces to:
\begin{equation}
\phi''-P^{2}\phi=0,
\end{equation}
So at the infinity $\phi$ behaves as
\begin{equation}\label{sol-infty}
\phi|_{r\to \infty} = B_1e^{r/P} + B_2 e^{-r/P}~,
\end{equation}
where $B$ are another integration constants. So if we assume that $B_1 = 0$ the function $\phi$ will be sharply decreasing function from the location of the brane, i.e. scalar field $\Phi$ will be localized on the brane.

We have demonstrate pure gravitational localization of scalar zero mode particles on the brane using approximate solutions (\ref{sol-0}) and (\ref{sol-infty}). We can obtain also numerical solutions to the equations (\ref{system_KG}). We use Runge-Couta 4-5 Scheme, with $10^{-11}$ error tolerance.
\\\\
Figure 2 displays the behaviour of the function $\nu (r)$ and its first derivative close to the brane.
\\\\

%%%%%%%%%%%%%%%%%%%%%%%%%%%%%%%%%%%%%%%%%%%%%
\begin{figure}[!htb]
\begin{center}
\includegraphics[width=0.6\textwidth]{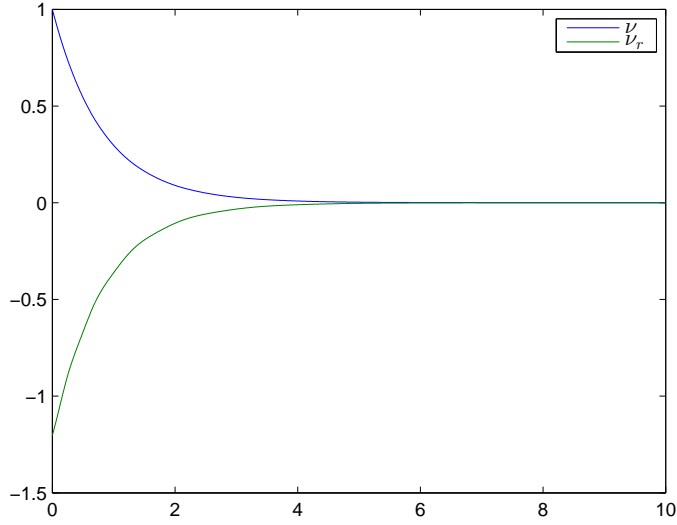}
\caption{Numerical solutions for $\nu$ and $\nu'$.}
\label{fig:sol}
\end{center}
\end{figure}
%%%%%%%%%%%%%%%%%%%%%%%%%%%%%%%%%%%%%%%%%%%

To proof localization of $\Phi$ we need to show that all integrals over $r$ in its action (\ref{Action_last}) is finite. On Figure 3 we display all $r$-depended factors in (\ref{Action_last}) which are multiplied by $r$. We see that they all are decreasing functions, i.e. (\ref{Action_last}) is integrable over $r$ and scalar field zero modes are localized on the brane.

%%%%%%%%%%%%%%%%%%%%%%%%%%%%%%%%%%%%%%%%%%%%%%%
\begin{figure}[!htb]
\begin{center}
\includegraphics[width=0.6\textwidth]{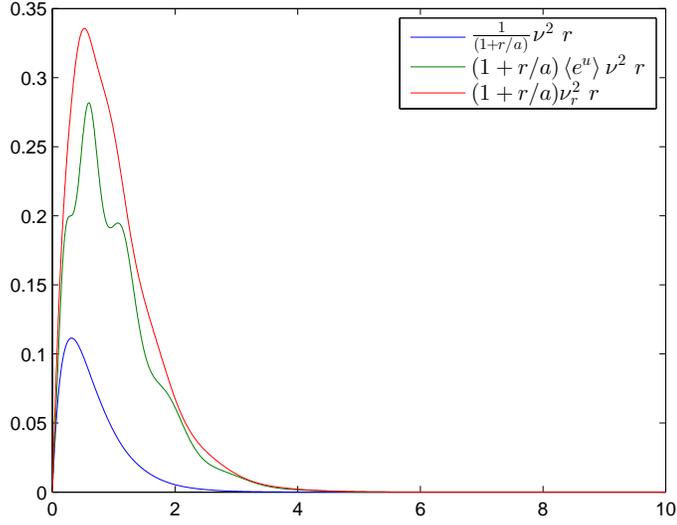}
\caption{$r$-depended factors in (\ref{Action_last}) multiplied by $r$}
\label{fig:int}
\end{center}
\end{figure}
%%%%%%%%%%%%%%%%%%%%%%%%%%%%%%%%%%%%%%%%%%%%

%%%%%%%%%%%%%%%%%%%%%%%%%%%%%%%%%%%%%%%%%%%%%%%%%%%%%%%%%%%%%%%%%%%%%%%%%%%%%%%%%%%%%%%

\section{Conclusions}

In this letter we have demonstrated the pure gravitational localization of scala field zero modes within the 6D model of the standing wave braneworld \cite{Wave}. The model represents a single (1+3)-brane in six dimensional  space-time with one large (infinite) and one small (compact) space-like extra dimensions. The trapping is provided by the rapid oscillations of gravitational; and ghost-like scalar fields in the bulk. The main differences of our model from 5D case is that the plane of the brane is warped symmetrically.

%%%%%%%%%%%%%%%%%%%%%%%%%%%%%%%%%%%%%%%%%%%%%%%%%%%%%%%%%%%%%%%%%%%%%%%%%%%%%%%%%%%%

\section*{Acknowledgments}

This research was supported by the grant of Shota Rustaveli National Science Foundation $\#{\rm DI}/8/6-100/12$.

%%%%%%%%%%%%%%%%%%%%%%%%%%%%%%%%%%%%%%%%%%%%%%%%%%%%%%%%%%%%%%%%%%%%%%%%%%%%%%%%%%%%%

\end{document}